\documentclass[lettersize,journal]{IEEEtran}
\usepackage{amsmath,amsfonts}
\usepackage{algorithmic}
\usepackage{array}
\usepackage[caption=false,font=normalsize,labelfont=sf,textfont=sf]{subfig}
\usepackage{textcomp}
\usepackage{stfloats}
\usepackage{url}
\usepackage{verbatim}
\usepackage{graphicx}
\def\BibTeX{{\rm B\kern-.05em{\sc i\kern-.025em b}\kern-.08em
    T\kern-.1667em\lower.7ex\hbox{E}\kern-.125emX}}
\usepackage{balance}
\usepackage{subfloat}
\usepackage{color}
\begin{document}
\title{Rotatable Antenna-array-enhanced Direction-sensing for Low-altitude Communication Network: Method and Performance}
\author{Jinbing Jiang, Feng Shu, Bin Deng, Maolin Li, Jiatong Bai, Yan Wang, Cunhua Pan, Jiangzhou Wang
\thanks{This work was supported in part by the National Natural Science Foundation of China under Grant U22A2002, the Hainan Province Science and Technology Special Fund under Grant ZDYF2024GXJS292, in part by the National Key Research and Development Program of China under Grant 2023YFF0612900.(Corresponding author: Feng Shu).
	
Jinbing Jiang, Bin Deng, Maolin Li, Jiatong Bai and Yan Wang are with the School of Information and Communication Engineering, Hainan University, Haikou, 570228, China. (e-mail: jiangjinbing1125@163.com; d2696638525@126.com; limaolin0302@163.com; 18419229733@163.com; yanwang@hainanu.edu.cn).

Feng Shu is with the School of Information and Communication Engineering and Collaborative Innovation Center of Information Technology, Hainan University, Haikou 570228, China, and also with the School of Electronic and Optical Engineering, Nanjing University of Science and Technology, Nanjing 210094, China. (e-mail: shufeng0101@163.com).

Cunhua Pan is School of Information Science and Engineering, Southeast University, Nanjing 210096, China. (Email:cpan@seu.edu.cn).

Jiangzhou Wang is School of Information Science and Engineering, Southeast University, Nanjing 210096, China and with the School of Engineering, University of Kent, Canterbury CT2 7NT, U.K. (Email:j.wang@seu.edu.cn).}

}

\markboth{Journal of \LaTeX\ Class Files,~Vol.~18, No.~9, September~2020}%
{How to Use the IEEEtran \LaTeX \ Templates}

\maketitle

\begin{abstract}
In a practical multi-antenna receiver, each element of the receive antenna array has a directive antenna pattern, which is still not fully explored and investigated in academia and industry until now. When the emitter is deviated greatly from the normal direction of antenna element or is close to the null-point direction, the sensing energy by array will be seriously attenuated such that the direction-sensing performance is degraded significantly, which will be faced in the low-altitude wireless network with a large amount of unmanned aerial vehicles (UAVs) in the coming future. To address such an issue, a rotatable array system is established with the directive antenna pattern of each element taken into account, where each element has the same antenna pattern. Then, the corresponding  the Cramer-Rao lower bound (CRLB)  is derived. Finally, a recursive rotation Root-MUSIC (RR-Root-MUSIC) direction-sensing method is proposed and its root-mean-squared error (RMSE) performance is evaluated by the derived CRLB. Simulation results show that the proposed rotation method converges rapidly with about ten iterations, and  make a significant enhancement on the direction-sensing accuracy in terms of RMSE when the target direction departs seriously far away from the normal vector of array. Compared with conventional Root-MUSIC, the sensing performance of the proposed RR-Root-MUSIC method is much closer to the CRLB.
\end{abstract}


\section{Introduction}
\IEEEPARstart{W}{ith} the rise of the low-altitude economy, unmanned aerial vehicles (UAVs) have gradually become an indispensable part of future wireless networks, enabling a wide range of applications from logistics distribution, emergency rescue to aerial inspection  \cite{Y. Wang}-\cite{J. Li}. To ensure the safe development of the low-altitude economy, a wireless communication network with high coverage and high reliability is of vital importance \cite{Z. Yan}. Right now cutting-edge technologies, such as directional modulation (DM) \cite{Wu X M}-\cite{X. Wu} and intelligent reflecting surface (IRS) aided communication \cite{Y. Teng}, offer effective approaches to enhance the communication performance, which is highly dependent on precise alignment between the transmit beam and the emitter. Therefore, it is of great significance to achieve high-precision sensing and tracking of UAVs with wide coverage areas.

Currently, the classical spatial spectral-based techniques, parametric-based estimation methods, the sparsity sensingbased techniques and deep learning-based methods are the main direction of arrival (DOA) estimation method \cite{DOA-MUSIC}. Based on these methods, the estimation performance has been significantly improved between computational efficiency and environmental adaptability. In terms of reducing system complexity and enhancing efficiency, a heterogeneous sub-connected hybrid analog–digital ($\mathrm{H^2AD} $) architecture was introduced in \cite{Hybrid} to address the limitation of time-efficiency. Based on this, the authors in \cite{Jiatong Bai} proposed a fully-digital $\mathrm{H^2AD} $ design and multi-modal learning strategy, which further reduced the complexity and enhanced the robustness of the system.  In addition, a fast ambiguous phase elimination method was proposed, which significantly reduced computational complexity and estimation time at the expense of a small performance cost in \cite{Shi B}. In terms of enhancing robustness in complex environments, aiming at the common mutual coupling effect and position error problems in UAV communication, a novel two-dimensional DOA auxiliary framework was proposed in \cite{F. Wen}, for anonymous UAV localization to effectively suppressing the influence of non-ideal factors on estimation accuracy. Although the above methods can approach the performance limit under ideal conditions, their effectiveness is highly dependent on the received signal-to-noise ratio (SNR). In practice, each element of the receive antenna array has a directive antenna pattern \cite{Q. Wu}. The current cellular network infrastructure is mainly planned and deployed for Terrestrial Users. To optimize ground coverage and suppress cell interference, existing base station (BS) antennas are usually equipped with a preset electrical/mechanical downtilt design \cite{X. Lin}. The traditional DOA estimation method is based on fixed antennas, it is difficult to ensure the validity of DOA estimation in the extreme scenarios.

To address this issue, some promising technologies, by designing more advanced beam patterns to enhance the accuracy of DOA estimation, such as the fluid antenna (FA) \cite{K.-K. Wong}-\cite{W. K. New} and mobile antenna (MA) \cite{L. Zhu}-\cite{X. Shao} architectures have been proven to increase system degrees of freedoms. Different from conventional fixed antenna arrays, the antenna elements can be flexibly reconfigured by FA/MA systems within a limited space, making full use of fluctuations in the spatial electromagnetic field to enhance communication or sensing performance \cite{M. Li}-\cite{Y. Si}. This characteristic is particularly significant in multipath scattering environments, and through position optimization, it can effectively avoid deep fading and maximize channel capacity. Existing research has made progress in many directions. Existing research has been advanced in several directions, including Bayesian reconstruction of FA channel estimation \cite{Z. Zhang}, FA-enabled DOA estimation under hybrid architecture \cite{J. Ren}, tensor-decomposition-based DOA estimation for MA-enabled MIMO \cite{R. Zhang}, and six-dimensional movable-antenna sensing scheme \cite{R. Schober}.

However, due to the ``ground-centered'' design paradigm, the low-altitude airspace often falls into the sub-lobe or zero lobe of the BS antenna pattern. This misalignment leads to a sharp drop in the signal-to-noise ratio of the high-altitude UAV reception, causing the DOA estimation error to exceed the acceptable range and rendering the sensing function ineffective. Although the FA/MA architecture can fine-tune the antenna position to mitigate fading, they essentially lack DoFs to redirect the main beam. Therefore, they are not sufficient to fundamentally address the challenge of "coverage blind spots" caused by angle mismatch.

To overcome the gain attenuation problem caused by directive antennas, the rotatable antenna (RA) architecture has proposed by \cite{T. Ma}. By mechanically adjusting the 3D orientation/boresight of each directive element while keeping its physical location fixed, RAs can reconfigure the effective array manifold without requiring element repositioning. 
This architecture is used to enhance the multi-user signal-to-noise ratio \cite{ICC}, increase the millimeter-wave coverage range of unmanned aerial vehicles \cite{X. Zhang}, and strengthen the physical layer security \cite{L. Dai}. Inspired by this, an iterative rotatable array system is proposed and applied to DOA estimation to enhance the coverage area of UAV communication systems. Our main contributions are summarized as follows:
\begin{enumerate}
	\item  To overcome the serious receive energy attenuation caused by the  antenna element directive pattern, a rotatable planar array system is established to enhance the sensing performance where each element is assumed to have the same antenna pattern. Different from the conventional fixed array system, the proposed system may near the emitter direction by multiple operations of sensing and rotation to improve the receive SNR. As SNR grows, the sensing accuracy will be improved correspondingly. The multiple sensing and rotation framework will be very suitable for the future low-altitude wireless networks with rich UAVs.

	\item  To achieve an excellent direction-sensing performance, a recursive rotation Root-MUSIC direction-sensing method, called RR-Root-MUSIC, is proposed. First, the method can sense the emitter direction, subsequently rotate the array boresight orientation to the estimated direction, sense again, rotate, etc. The about sensing and rotation loop is repeated until the difference between the adjacent two sensing values is below the predefined threshold. To evaluate the performance of the proposed RR-Root-MUSIC method, the closed-form expression of the corresponding  the Cramer-Rao lower bound (CRLB), considering the directive antenna element is derived. The derived CRLB will be used as a benchmark in the following.

	\item  	  Simulation is conducted to assess the error performance of the proposed RR-Root-MUSIC. In accordance with simulation results, the proposed RR-Root-MUSIC performs much better than conventional fixed Root-MUSIC and is closer to the corresponding CRLB. More important, the proposed method  has a fast convergent speed of about ten iterations.  When the target direction departs far away from the normal vector of array, the performance gain achieved by recursive sensing and rotation process is tremendous.  For example, at $\theta\le 15^\circ$, where $\theta$ denotes the direction difference between the target direction and the  antenna array plane, the achievable performance gain of the proposed RR-Root-MUSIC over conventional Root-MUSIC will be up to  about three-order magnitude or more.  In particular, as $\theta$  nears the array plane, i.e. far away from the normal direction of rotation array, the achievable performance gain will grow gradually and become more significant.

\end{enumerate}

The remainder of this research is organized as follows. The system model of rotatable array is described in Section II. In Section III, the RR-Root-MUSIC is proposed. In Section IV, corresponding CRLB is also derived. Moreover, section V presents the experimental results, with conclusions provided in Section VII.

\section{System Model}
\noindent
\begin{figure}
	\centering
	\includegraphics[width=0.7\linewidth]{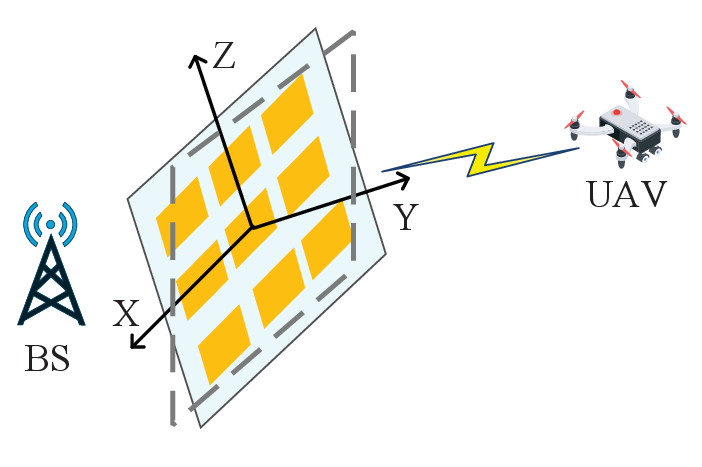}
	\caption{Rotatable array system for low-altitude communication network.}
	\label{fig:ra}
\end{figure}
As shown in Fig. \ref{fig:ra}, a rotatable array system for low-altitude communication network is considered. Different from conventional fixed array, a uniform planar array (UPA) with $M\times N$ directive antenna elements is deployed on the $x$-$z$ plane, where $N$ and $M$ elements are arranged along the $x$-axis and $z$-axis respectively. The antenna spacing in two axes are denoted by $d_x$ and $d_z$, respectively. The geometric reference point of the array is taken as its geometric center. Furthermore, the location of the $(n,m)$-th antenna is 
\begin{align}
	\mathbf{p}_{n,m}=[x_n,0,z_m  ]^T,
\end{align}	
where $x_n =(n-\frac{N-1}{2} ) d_x$, $z_m=(m-\frac{M-1}{2})d_z$,  $n=0,...,N-1$ and $m=0,...,M-1$. 

\begin{figure}[!t]
	\centering
	\subfloat[Initial orientation]{
		\includegraphics[width=0.2\textwidth]{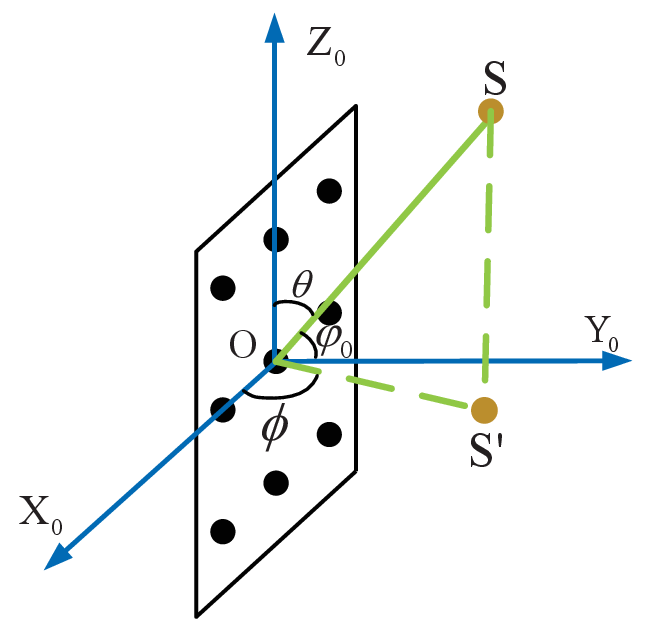}%
		\label{subfig:xyz-a}
	}
	\hfil
	\subfloat[Rotated orientation]{
		\includegraphics[width=0.2\textwidth]{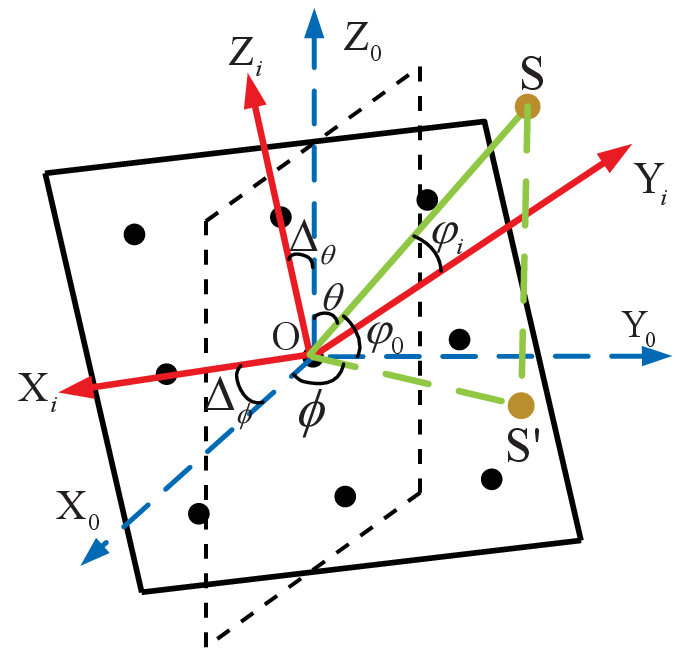}%
		\label{subfig:xyz-b}
	}
	\caption{Illustration of the geometric relationship for the rotatable array and emitter.}
	\label{fig:xyz}
\end{figure}

Fig.~\ref{fig:xyz} illustrates the geometric configuration of the rotatable antenna array. Let the elevation angle $\theta$ denote the angle between emitter direction and array plane, and the azimuth angle $\phi$ represent the angle between the projection of the emitter direction on the $x$-$y$ plane and the array plane. Accordingly, the unit direction vector of the emitter can be expressed as
\begin{align}\label{OS}
	\overrightarrow{OS_0}(\theta,\phi) &=[\sin\theta\cos\phi,\sin\theta\sin\phi,\cos\theta]^T.
\end{align}

Define a narrowband signal $s$ is radiated by the emitter that satisfies
$\mathbb{E}\{|s|^{2}\}=P_{t}$. As shown in Fig. \ref{fig:xyz}\subref{subfig:xyz-a}, a boresight deflection angle $\varphi_{0}$ is formed between the emitter direction and the normal direction of the array. The corresponding directive antenna gain is denoted by $g(\varphi_{0})$. Thus, the received signal by the $(n,m)$-th antenna element can be written as 
\begin{align}\label{y0}
	y_{n,m}&=sg(\varphi_0)e^{j\frac{2\pi}{\lambda}( x_n \sin\theta \cos\phi+z_m\cos\theta)} +n_{n,m},
\end{align}
where $\lambda$ is the wavelength and $n_{n,m}$ denotes the additive white Gaussian noise (AWGN). 

Let us define that $\mathbf{y}\in \mathbb{ C}^{1\times NM}$ and $\mathbf{n}\in \mathbb{ C}^{1\times NM}$ represent the received data matrix and the AWGN noise matrix, respectively. Assume the mean value of AWGN noise is zero and the root-mean-squared error (RMSE) is $\sigma^2 $.  Then, $\mathbf{y}$ can be repressed as
\begin{equation}
	\mathbf{y}=s\,g(\varphi_0)\,\mathbf{a}_z(\theta)\otimes  \mathbf{a}_x^H(\theta, \phi) +\mathbf{n},
\end{equation}
where $\otimes$ is the Kronecker product and the so-called array manifold defined by
\begin{equation}
	\mathbf{a}_x(\theta, \phi) = 
	\begin{bmatrix}
		e^{j\frac{2\pi}{\lambda} x_0  \sin\theta\cos\phi} \\
		\cdots  \\
		e^{j\frac{2\pi}{\lambda}x_n \sin\theta\cos\phi}  \\
		\cdots  \\
		e^{j\frac{2\pi}{\lambda} x_{N-1}  \sin\theta\cos\phi}
	\end{bmatrix},
\end{equation}
and
\begin{equation}
	\mathbf{a}_z(\theta) = \begin{bmatrix}
		e^{j\frac{2\pi}{\lambda} z_0 \cos\theta} 
		\cdots 
		e^{j\frac{2\pi}{\lambda} z_m \cos\theta} 
		\cdots 
		e^{j\frac{2\pi}{\lambda} z_{M-1} \cos\theta}
	\end{bmatrix}^H.
\end{equation}

%

In Fig. \ref{fig:xyz}\subref{subfig:xyz-a}, the rotation array is placed on the $x–z$ plane at initially. Hence, the normal vector of array is aligned with the $y$-axis and can be expressed as
\begin{align}
	\overrightarrow{OY_0}=[0,\,1,\,0]^T.
\end{align}

According to the well-known cosine identity \cite{oxford}
\begin{align}\label{cos}
	\cos\theta =\frac{\left \langle \mathbf{a},\mathbf{b}  \right \rangle}{|\mathbf{a}||\mathbf{b}|},
\end{align}
the boresight deflection angle $\varphi_{0}$ can be calculated as
\begin{align}
	\varphi_0=\arccos(\sin\theta\sin\phi).
\end{align}

As illustrated in Fig. \ref{fig:xyz}\subref{subfig:xyz-b}, the array is rotated counterclockwise about the $x$-axis by $\Delta_{\theta}$ and clockwise about the $z$-axis by $\Delta_{\phi}$. The coordinate system is rotated together with the array. Correspondingly, the new emitter direction vector is represented as
\begin{align}
	\overrightarrow{OS_i}=\begin{bmatrix}\sin\theta\cos(\Delta_{\phi}+\phi)
		\\\sin\Delta_{\theta}\cos\theta+\sin\theta\sin(\Delta_{\phi}+\phi)\cos\Delta_{\theta}
		\\-\sin\Delta_{\theta}\sin\theta\sin(\Delta_{\phi}+\phi)+\cos\theta\cos\Delta_{\theta}
	\end{bmatrix},
\end{align}
in accordance with (\ref{cos}), the equation can be obtained as follows
\begin{align}
	\cos\varphi_i =&\frac{\left \langle\overrightarrow{OS_i},\overrightarrow{OY}\right \rangle}{|\overrightarrow{OS_i}||\overrightarrow{OY}|}\nonumber\\
	=&\sin\theta\cos\Delta_{\theta}\sin(\phi+\Delta_{\phi})+\cos\theta\sin\Delta_{\theta},
\end{align}
hence, the new deflection angle $\varphi_i$ is 
\begin{align}\label{phi}
	&\varphi_i(\Delta_{\theta},\Delta_{\phi},\theta,\phi)\nonumber\\
	=&\arccos(\sin\theta\cos\Delta_{\theta}\sin(\phi+\Delta_{\phi})+\cos\theta\sin\Delta_{\theta}). 
\end{align}

After rotation, the new received signal at the $(n,m)$-th antenna becomes 
\begin{equation}\label{y}
	\tilde{y}_{n,m}=s\,g(\varphi)\,e^{j\psi_{n,m}}+n_{n,m},
\end{equation}
where the phase term is given by
\begin{align}\label{psi}
	\psi &_{n,m}(\Delta_{\theta},\Delta_{\phi},\theta,\phi)
	=\frac{2\pi}{\lambda}\Big[
	x_{n}\sin\theta\cos(\phi+\Delta_{\phi})
	\nonumber\\
	&+\,z_{m}\big(
	\cos\theta\cos\Delta_{\theta}
	-\sin\theta\sin\Delta_{\theta}\sin(\phi+\Delta_{\phi})
	\big)
	\Big].
\end{align}

Correspondingly the new received signal matrix $\tilde{\mathbf{y}}$ is expressed as
\begin{align}	
	\tilde{\mathbf{y}}
	=sg(\varphi)\tilde{\mathbf{a}}_z(\theta)\otimes  \tilde{\mathbf{a}}_x^H(\theta, \phi)  +\mathbf{n},
\end{align}
with the rotated array response vectors as
\begin{align}
\tilde{\mathbf{a}}_z(\theta)=\begin{bmatrix}
	e^{j\frac{2\pi}{\lambda}x_0 \sin\theta\cos(\phi+\Delta_{\phi})} \\
	\vdots \\
	e^{j\frac{2\pi}{\lambda}x_n \sin\theta\cos(\phi+\Delta_{\phi})} \\
	\vdots \\
	e^{j\frac{2\pi}{\lambda}x_{N-1} \sin\theta\cos(\phi+\Delta_{\phi})}
\end{bmatrix},
\end{align}
and
\begin{align}
	\tilde{\mathbf{a}}_x(\theta, \phi)=\begin{bmatrix}
		e^{j\frac{2\pi}{\lambda} z_0 (\cos\theta\cos\Delta_{\theta}-\sin\theta\sin\Delta_{\theta}\sin(\phi+\Delta_{\phi}))} \\
		\vdots \\
		e^{j\frac{2\pi}{\lambda} z_m (\cos\theta\cos\Delta_{\theta}-\sin\theta\sin\Delta_{\theta}\sin(\phi+\Delta_{\phi}))} \\
		\vdots \\
		e^{j\frac{2\pi}{\lambda} z_{M-1} (\cos\theta\cos\Delta_{\theta}-\sin\theta\sin\Delta_{\theta}\sin(\phi+\Delta_{\phi}))}
	\end{bmatrix}.
\end{align}

\section{Proposed Recursive Rotation Root-MUSIC (RR-Root-MUSIC) Method}
\noindent
\begin{figure}[t]
	\centering
	\subfloat[Initial]{
		\includegraphics[width=0.13\textwidth]{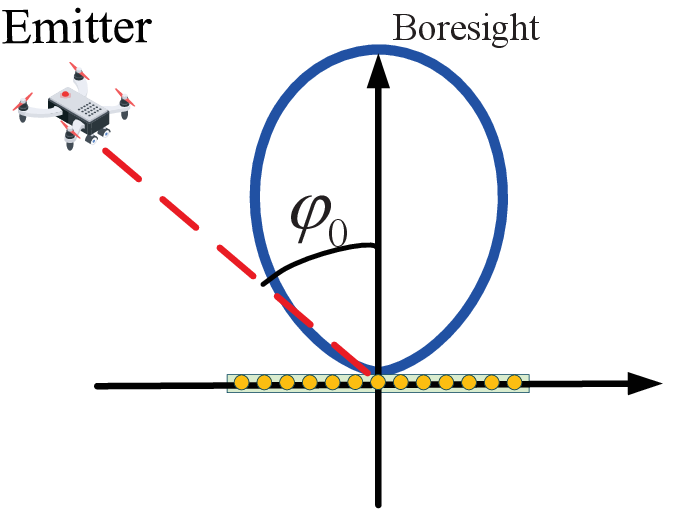}
		\label{a}
	}
	\hfill
	\subfloat[Rotating]{
		\includegraphics[width=0.13\textwidth]{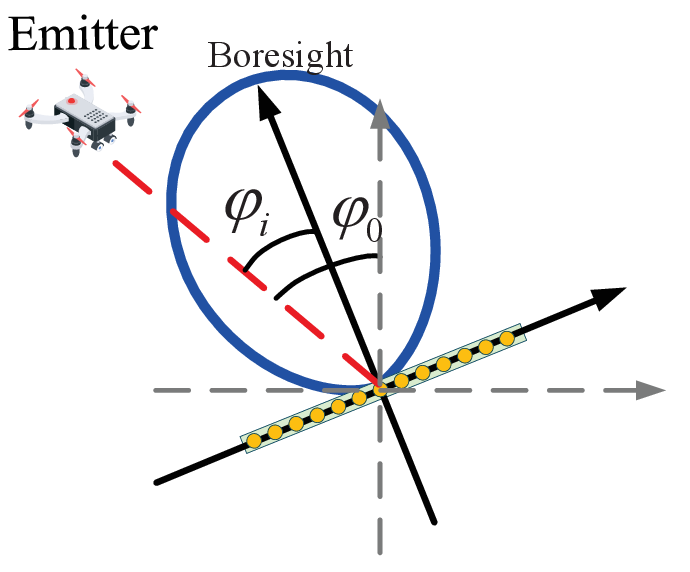}
		\label{b}
	}
	\hfill
	\subfloat[Convergent]{
		\includegraphics[width=0.13\textwidth]{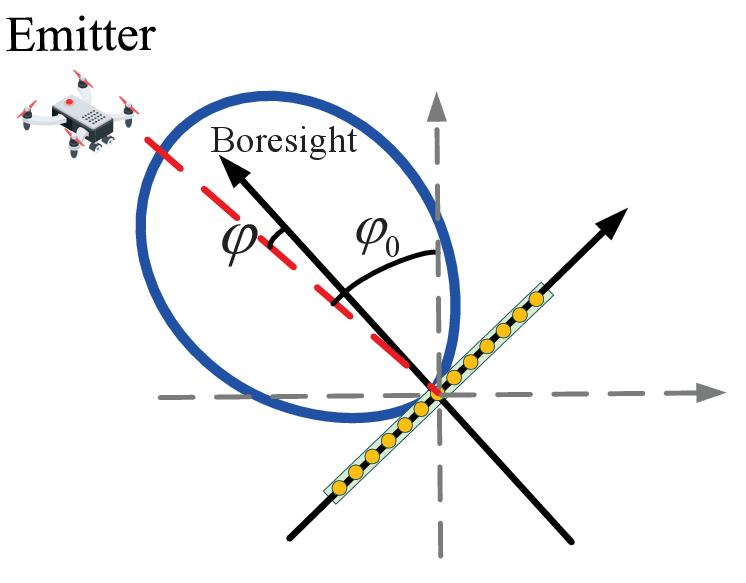}
		\label{c}
	}
	\caption{Illustration of boresight deflection angle and directive array pattern.}
	\label{abc}
\end{figure}
Due to directive antenna patterns, the sensing power is significantly attenuated when the emitter direction deviates from the normal vector. Therefore, in this section, a RR-Root-MUSIC method is proposed. As shown in Fig. \ref{abc}\subref{a}, when the emitter direction is far away from the normal vector of array, Root-MUSIC method is applied at the initial orientation which provides a coarse estimation value at the lower directional gain. According to this estimation angle, the array is rotated into the main-lobe region as Fig.~\ref{abc}\subref{b}, which increases the sensing power and improves the estimation condition. With the updated orientation, a new estimation angle is obtained, and the process repeats. As shown in Fig.~\ref{abc}\subref{c}, the offset angle gradually shrinks over iterations, leading to higher effective SNR and eventual convergence toward the lower RMSE.

At first, the Root-MUSIC algorithm is used to estimate the elevation and azimuth $(\theta,\phi)$ of initial orientation array. Assume that $\tilde{\mathbf{y}}_(k)$ represents the received signal vector at the rotation array for the $k$-th snapshot.
The sample covariance matrix $\mathbf{\hat{R}}$ of $\tilde{\mathbf{y}}(k)$ is calculated as 
\begin{equation}\label{R}
	\mathbf{\hat{R}}=\frac{1}{K}\sum_{k=0}^{K}\tilde{\mathbf{y}}(k)\tilde{\mathbf{y}}^H(k),
\end{equation}
then, through eigenvalue decomposition, Eq. (\ref{R}) is decomposed into	
\begin{align}	
	\mathbf{\hat{R}}=\mathbf{U}\Lambda \mathbf{U}^H=[\mathbf{U}_{s} \mathbf{U}_{n}]\sum[\mathbf{U}_{s}\mathbf{U}_{n}]^H,
\end{align}
where $\mathbf{U}_{s}=[\mathbf{u}_1]$ is the signal subspace, and $\mathbf{U}_{n}=\left[\mathbf{u}_2, \mathbf{u}_3,...,\mathbf{u}_{MN}\right]$ is the noise subspace. According to \cite{DOA-MUSIC}, the MUSIC spectrum of Eq. (\ref{y}) is given by	
\begin{align}	
	P(\theta,\phi)&=\frac{1}{\mathbf{a}^H(\theta,\phi)\mathbf{U}_{n}\mathbf{U}_{n}^H\mathbf{a}(\theta,\phi)}\nonumber\\&=\frac{1}{\sum_{j=2}^{MN}|\mathbf{a}^H(\theta,\phi)\mathbf{u}_j|^2},
\end{align}	
which has the peak corresponding to the desired DOA estimation.	

The Root-MUSIC method is reliable due to its excellent asymptotic performance of achieving the CRLB. Thus, the Root-MUSIC is adopted to generate candidate solutions. The array manifold $\mathbf{a}(\theta, \phi)$ can be further expressed as
\begin{align}
	\mathbf{a}(z_x, z_z)&=\mathbf{a}(z_x)\otimes\mathbf{a}( z_z)\nonumber\\&=\begin{bmatrix}z_x^{-(N-1)/2}
		\\z_x^{-(N-3)/2}
		\\\dots
		\\z_x^{(N-1)/2}
	\end{bmatrix} \otimes 
	\begin{bmatrix}z_z^{-(M-1)/2}
		\\z_z^{-(M-3)/2}
		\\\dots
		\\z_z^{(M-1)/2}
	\end{bmatrix},
\end{align}	
where 
\begin{align}	
	z_x=e^{j\frac{2\pi}{\lambda}d_x\sin\theta \cos\phi},
	z_z=e^{j\frac{2\pi}{\lambda}d_z\cos\theta}.
\end{align}	

Then, the polynomial equation is	
\begin{align}\label{fz}	
	f(z_x, z_z)=\mathbf{a}^H(z_x, z_z)\mathbf{U}_{ni}\mathbf{U}_{ni}^H\mathbf{a}(z_x, z_z),
\end{align}	
the objective of Root-MUSIC is to find the root pair $(z_x, z_z)$ on the unit circle that minimizes the function $f(z_x, z_z)$.

Therefore,  the feasible solutions of UPA is 
\begin{align}	
	\hat{\theta}=\arccos(\frac{\angle z_z }{2\pi d_z/\lambda}).
\end{align}

It is worth noting that the feasible solution of $\hat{\phi}$ cannot be directly calculated. Due to the phase ambiguity that needs to be eliminated, denoted as 
\begin{align}	
	\cos\hat{\phi}=\frac{\angle z_x }{2\pi d_z\sin\hat{\theta}/\lambda}.
\end{align}		
And $\cos\hat{\phi}$ has a period of 2$\pi$, expressed as
\begin{align}	
	\cos\hat{\phi}=\cos(2\pi-\hat{\phi}).
\end{align}

The MUSIC pseudo-spectrum $P_{\text{MUSIC}}(\theta,\phi)$ is evaluated over a predefined search grid by testing the orthogonality between each candidate steering vector and the estimated noise subspace. In particular, the formula is used to measures the alignment between the array manifold and the noise subspace as follows
\begin{equation}
	\mathbf{a}^{H}(\hat{\theta},\hat{\phi})\mathbf{U}_{n}\mathbf{U}_{n}^{H}\mathbf{a}(\hat{\theta},\hat{\phi}).
\end{equation}

Finally, by searching for the grid point with the largest MUSIC hyperspectral value in the array, the estimated values of the non-rotation elevation angle and azimuth angle are obtained and expressed as $(\hat{\theta},\hat{\phi})$.

Next, based on the estimated value from the previous time, the array will be rotated. Without loss of generality, we present the procedure for the $(i + 1)$-th iteration rotation in the following, where $i=0,1,2,...,L$. 

Assume that the estimated elevation and azimuth at iteration $i$-th are $(\theta_i,\phi_i)$. Then, the rotation elevation and azimuth angles of $i$-th are 
\begin{align}	
	\Delta_{\theta_i}=90^o-\hat{\theta}_{i},\nonumber\\
	\Delta_{\phi_i}=90^o-\hat{\phi}_{i},
\end{align}
where the array is rotated as counterclockwise around the x-axis by $\Delta_{\theta_i}$, and clockwise along the z-axis by $\Delta_{\phi_i}$. In addition, the coordinate axes is rotated along with the array.

For the new coordinate axes and array, the corresponding true elevation and azimuth angles $(\theta_{i+1},\phi_{i+1})$ is
\begin{align}	
	\theta_{i+1}=\theta+\Delta_{\theta_i},\nonumber\\
	\phi_{i+1}=\phi+\Delta_{\phi_i},
\end{align}
and the new normal vector $\overrightarrow{OS}(\theta_{i+1},\phi_{i+1})$, which is the true normal vector after rotation, can be obtained as
\begin{align}\label{Ri}
	\overrightarrow{OS}(\theta_{i+1},\phi_{i+1}) =\mathbf{R}_i^H\cdot\overrightarrow{OS}(\theta,\phi),
\end{align}	
where $\mathbf{R}_i$ is the $i$-th rotation matrix. It can be written as
\begin{align}
	\mathbf{R}_i&=\mathbf{R}_{zi}\mathbf{R}_{xi}\\&=\begin{bmatrix}
		\cos\Delta_{\phi_i}&  -\sin\Delta_{\phi_i}& 0\\
		\sin\Delta_{\phi_i}&  \cos\Delta_{\phi_i}& 0\\
		0&  0& 1
	\end{bmatrix}\cdot\begin{bmatrix}
		1&	0&  0 \\
		0	& 	\cos\Delta_{\theta_i}&  \sin\Delta_{\theta_i}\\
		0	& 	-\sin\Delta_{\theta_i}&  \cos\Delta_{\theta_i}
	\end{bmatrix}.\nonumber
\end{align}	

According to Eq. (\ref*{phi}), the new emitter direction after rotation is closer to the normal vector of the array, i.e., the higher SNR can be achieved. For the new emitter direction, a new estimated angle $(\hat{\theta}_{i+1},\hat{\phi}_{i+1})$ can be obtained via Root-MUSIC method.


Next, restore the coordinate axes to their orientation before rotation. Corresponding to the estimated angle $(\hat{\theta}_{i+1},\hat{\phi}_{i+1})$, the estimated vector of the emitter direction after rotation is represented as
\begin{align}
	&\overrightarrow{OS}(\hat{\theta}_{i+1},\hat{\phi}_{i+1})\nonumber\\ =&[\sin\hat{\theta}_{i+1}\cos\hat{\phi}_{i+1},\sin\hat{\theta}_{i+1}\sin\hat{\phi}_{i+1},\cos\hat{\theta}_{i+1}]^T.
\end{align}

According to Eq. (\ref{Ri}), the restore incident direction  $\overrightarrow{OS}'(\hat{\theta}_{i+1},\hat{\phi}_{i+1})$ can be calculated through the inverse matrix of the rotation matrix as
\begin{align}
	\overrightarrow{OS}'(\hat{\theta}_{i+1},\hat{\phi}_{i+1}) &=(\mathbf{R}^H)^{-1}\cdot\overrightarrow{OS}(\hat{\theta}_{i+1},\hat{\phi}_{i+1}).
\end{align}

Therefore, the ($i$+1)-th estimated elevation and azimuth angles $(\bar{ \theta}_{i+1},\bar{\phi} _{i+1})$ are
\begin{align}
	\bar{ \theta}_{i+1}&=\arccos(\overrightarrow{OS}'(3)),\nonumber\\
	\bar{\phi}_{i+1}&=\arctan(\overrightarrow{OS}'(2),\overrightarrow{OS}'(1)).
\end{align}

Define that $\epsilon_e$ is an estimation error threshold, expressed as
\begin{align}\label{epsilon}
	|\bar{ \theta}_{i+1}-\bar{ \theta}_{i}| \le \epsilon_e,\nonumber\\
	|\bar{ \phi}_{i+1}-\bar{ \phi}_{i}| \le \epsilon_e,
\end{align}

where $\epsilon_e=0.01^\circ$, when meets Eq. (\ref{epsilon}), the iteration process is stopped. And the final estimated elevation and azimuth angles are $(\bar{ \theta}_{i+1},\bar{ \phi}_{i+1})$. Otherwise, repeat the above steps.

\section{Derived CRLB for Rotatable Array}
\noindent

The CRLB provides a benchmark for the minimum achievable RMSE of any unbiased estimator. According to literature \cite{A. Wang}, \cite{S. Kay} and \cite{T. Engin}, to evaluate the performance of the proposed RR-Root-MUSIC method, the CRLB of the rotation array is derived. The derivation process is as follows. 

Based on the received signal model in Eq. (\ref{y}), the sensing vector at the $(n,m)$-th antenna element can be rewritten as
\begin{align}
	\tilde{y}_{n,m}=\tilde{\mu}_{n,m}+n_{n,m},
\end{align}
where 
\begin{align}\label{mu}
	\tilde{\mu}_{n,m}=s\,g(\varphi)\,e^{j\psi _{n,m}(\Delta_{\theta},\Delta_{\phi},\theta,\phi)}.
\end{align}

Donate the received data matrix as  
\begin{align}
\tilde{\mathbf{y}} = 
\big[
\tilde{y}_{0,0},\, \tilde{y}_{0,1},\dots,
\tilde{y}_{M-1,N-1}
\big]^{T},
\end{align}
and define the parameter vector as follows
\begin{align}
\boldsymbol{\theta}=[\theta,\phi]^{T}.
\end{align}

Referring to \cite{T. Engin}, the likelihood function of $\tilde{\mathbf{y}}$ is given by 
\begin{align}
	p(\tilde{\mathbf{y}};\boldsymbol{\theta})
	=(\pi\sigma^{2})^{-MN}
	\exp\!\left(
	-\frac{\|\tilde{\mathbf{y}}
		-\boldsymbol{\mu}(\boldsymbol{\theta})\|^{2}}
	{\sigma^{2}}
	\right),
\end{align}
where $\boldsymbol{\mu}(\boldsymbol{\theta})$ is the $\tilde{\mu}_{n,m}$ terms in the same order as $\tilde{\mathbf{y}}$.

In order to derive the joint CRLB for the elevation and azimuth angles, the Fisher information matrix (FIM) \cite{S. Kay} is introduced as
\begin{equation}\label{I}	
	[\mathbf{F}(\boldsymbol{\theta })]_{ij}=-E\left [\frac{\partial ^2\ln p(\tilde{\mathbf{y}};\boldsymbol{\theta } )}{\partial\theta_i\partial\theta_j }\right ].
\end{equation}

Further, Eq. (\ref{I}) can be simplified as
\begin{align}	
	&[\mathbf{F}(\boldsymbol{\theta })]_{ij}=\frac{2}{\sigma^2}\sum_{n=0}^{N-1}\sum_{m=0}^{M-1}\Re\{\frac{\partial \tilde{\mu}_{n,m} }{\partial\theta_i}\cdot \frac{\partial \tilde{\mu}^*_{n,m} }{\partial\theta_j}\},
\end{align}
where 
\begin{align}
	&\Re\{\frac{\partial \tilde{\mu}_{n,m} }{\partial\theta_i}\cdot \frac{\partial \tilde{\mu}^*_{n,m} }{\partial\theta_j}\}\nonumber\\
	=&s^2\begin{bmatrix}
	(\frac{\partial g }{\partial\theta})^2+ g^2 (\frac{\partial \psi }{\partial\theta})^2& \frac{\partial g }{\partial\theta}\frac{\partial g }{\partial\phi}+g^2 \frac{\partial \psi}{\partial \theta}\frac{\partial \psi}{\partial \phi}\\
	\frac{\partial g }{\partial\theta}\frac{\partial g }{\partial\phi}+g^2 \frac{\partial \psi}{\partial \theta}\frac{\partial \psi}{\partial \phi}	&(\frac{\partial g }{\partial\phi})^2+ g^2 (\frac{\partial \psi }{\partial\phi})^2
\end{bmatrix},
\end{align}
and $\psi=\psi_{n,m}$, $g=g(\varphi)$.

In accordance with Eq. (\ref{psi}), by taking the partial derivative of $\psi_{n,m}$, we have
\begin{multline}	
	\frac{\partial \psi_{n,m}}{\partial \theta} = \frac{2\pi}{\lambda} \bigl[ x_n\cos\theta\cos(\phi+\Delta_{\phi}) \\
	- z_m(\sin\theta\cos\Delta_{\theta} + \cos\theta\sin\Delta_{\theta}\sin(\phi+\Delta_{\phi})) \bigr],
\end{multline}
and
\begin{align}	
	\frac{\partial \psi_{n,m} }{\partial \phi}=-\frac{2\pi}{\lambda}\sin\theta[&x_n\sin(\phi+\Delta_{\phi})\nonumber\\&+z_m\sin\Delta_{\theta}\cos(\phi+\Delta_{\phi}))].
\end{align}

Assume that $N$ and $M$ are odd numbers and the array is symmetrical. Then there are $\sum_{n,m}x_n=0$, $\sum_{n,m}z_m=0$, $\sum_{m,n}x_n\,z_m=0$. Let us define
\begin{align}	
	S_n&=\sum^{N-1}_{n=0}x_n^2=\frac{N(N^2-1)}{12}d_x^2,\nonumber\\
	S_m&=\sum^{M-1}_{m=0}z_m^2=\frac{M(M^2-1)}{12}d_z^2.
\end{align}

Based on the above results, the single-snapshot FIM for the DOA estimation of the rotation array can be obtained as 
\begin{align}
	\mathbf{F}_{\text{(1 snap)}}(\boldsymbol{\theta})
	&= \frac{2s^{2}}{\sigma^{2}}
	\begin{bmatrix}
		P & Q\\
		Q & R
	\end{bmatrix},
\end{align}
where
\begin{align}\label{PQR}	
	P &= MN\left(\frac{\partial g}{\partial\theta}\right)^2 + d_1g^2\left[\cos^2\theta\cos^2(\phi+\Delta_{\phi})MS_n\right. \nonumber \\
	&\quad \left. + (\cos\theta\sin\Delta_{\theta}\sin(\phi+\Delta_{\phi}) + \sin\theta\cos\Delta_{\theta})^2 NS_m\right], \nonumber \\
	Q &= MN\left(\frac{\partial g}{\partial\theta}\frac{\partial g}{\partial\varphi}\right) + d_1g^2\left[-\sin\theta\cos\theta\cos(\phi+\Delta_{\phi})\right. \nonumber \\
	&\quad \left. \times \sin(\phi+\Delta_{\phi})(MS_n - NS_m\sin^2\Delta_{\theta}) \right. \nonumber \\
	&\quad \left. + \sin\Delta_{\theta}\cos\Delta_{\theta}\sin^2\theta\cos(\phi+\Delta_{\phi})NS_m\right], \nonumber \\
	R &= MN\left(\frac{\partial g}{\partial\varphi}\right)^2 + d_1g^2\sin^2\theta\left[\sin^2(\phi+\Delta_{\phi})MS_n\right. \nonumber \\
	&\quad \left. + \sin^2\Delta_{\theta}\cos^2(\phi+\Delta_{\phi})NS_m\right],\nonumber \\
	&d_1=(\frac{2\pi}{\lambda})^2.
\end{align}

Then, the CRLB for $K$ independent snapshots can be calculated as
\begin{align}\label{I1}
	\mathbf{F}(\textcolor{blue}{\boldsymbol{\theta}})=\frac{2\,K\,s^2}{\sigma^2}\begin{bmatrix}
		P	& Q\\
		Q	& R
	\end{bmatrix}.
\end{align}

According to \cite{T. Engin}, a general antenna directional pattern is defined by
\begin{align}
	G(\varphi)=
	\begin{cases}\sqrt{G_0}
		\cos^{p}(\varphi), & \varphi\in(-\frac{\pi}{2},\frac{\pi}{2}) \\
		0, & \text{otherwise},
	\end{cases} 
\end{align}
where $G_0=2(2p+1)$ is the maximum gain in the normal direction (i.e., $\varphi$ = 0) that meets the law of power conservation.

In addition, the channel power gain between emitter and array can be modeled as  
\begin{align}\label{g}
	g(\varphi)=
	\begin{cases}\sqrt{\frac{A}{4\pi r^2 }G_0}\cos^{p}(\varphi), & \varphi\in(-\frac{\pi}{2},\frac{\pi}{2}) \\
		0, & \text{otherwise},
	\end{cases} 
\end{align}
where the integral space $A$ corresponds to the surface region of rotatable array and $r$ is the distance between emitter and array. To simplify, a constant is defined as $g_0 =\sqrt{\frac{A}{4\pi r^2 }G_0}$.

Hence, substituting Eq. (\ref{g}) into Eq. (\ref{phi}), it can be rewritten as
\begin{align}\label{gvarphi}
	g(\varphi)=&g(\varphi(\Delta_{\theta},\Delta_{\phi},\theta,\phi))\nonumber\\
	=&g_0(\sin\theta\cos\Delta_{\theta}\sin(\phi+\Delta_{\phi})+\cos\theta\sin\Delta_{\theta})^p\nonumber\\
	=&g_0\alpha^p,
\end{align}
where 
\begin{align}
	&\alpha(\Delta_{\theta},\Delta_{\phi},\theta, \phi )\nonumber\\
	=&\sin\theta\cos\Delta_{\theta}\sin(\phi+\Delta_{\phi})+\cos\theta\sin\Delta_{\theta}.
\end{align}

The derivatives of $\theta$ and $\phi$ of the function $g(\varphi)$ are given by
\begin{align}\label{g1}	
	\frac{\partial g}{\partial\theta }=g_0p \alpha^{p-1} \beta_1,
\end{align}
and
\begin{align}\label{g2}	
	\frac{\partial g}{\partial\phi }&=g_0p \alpha^{p-1} \beta_2 ,
\end{align}
where
\begin{align}
	&\beta_1 (\Delta_{\theta},\Delta_{\phi},\theta, \phi )\nonumber\\
	=&\cos\theta\cos\Delta_{\theta}\sin(\phi+\Delta_{\phi})-\sin\theta\sin\Delta_{\theta},
\end{align}
and
\begin{align}
	\beta_2 (\Delta_{\theta},\Delta_{\phi},\theta, \phi )=\sin\theta\cos\Delta_{\theta}\cos(\phi+\Delta_{\phi}).
\end{align}

According to Eq. (\ref{g1}) and Eq. (\ref{g2}), Eq. (\ref{PQR}) can be further expressed as
\begin{align}	
	P &= MN(g_0p\alpha^{p-1}\beta_1)^2 \nonumber \\
	&\quad + d_1(g_0\alpha^p)^2 [\cos^2\theta\cos^2(\phi+\Delta_{\phi})MS_n \nonumber \\
	&\quad + (\cos\theta\sin(\phi+\Delta_{\phi})\sin\Delta_{\theta} \nonumber \\
	&\quad + \sin\theta\cos\Delta_{\theta})^2 NS_m], \nonumber \\	
	Q &= MN(g_0p\alpha^{p-1})^2\beta_1\beta_2 \nonumber \\
	&\quad + d_1(g_0\alpha^p)^2 [-\sin\theta\sin(\phi+\Delta_{\phi}) \nonumber \\
	&\quad \times (MS_n - NS_m\sin^2\Delta_{\theta})\cos\theta\cos(\phi+\Delta_{\phi}) \nonumber \\
	&\quad + \sin\Delta_{\theta}\cos\Delta_{\theta}\sin^2\theta\cos(\phi+\Delta_{\phi})NS_m], \nonumber \\
	R &= MN(g_0p\alpha^{p-1}\beta_2)^2 \nonumber \\
	&\quad + d_1(g_0\alpha^p)^2 \sin^2\theta[\sin^2(\phi+\Delta_{\phi})MS_n \nonumber \\
	&\quad + \sin^2\Delta_{\theta}\cos^2(\phi+\Delta_{\phi})NS_m].
\end{align}

The CRLB can be given by
\begin{align}
	CRLB =\mathbf{F}^{-1}(\boldsymbol{\theta})=\frac{\sigma^2}{2Ks^2}\frac{1}{PR-Q^2}\begin{bmatrix}
		R	& -Q\\
		-Q	& P
	\end{bmatrix} .
\end{align}

Therefore, the lower bound of the RMSE of the $\hat{\theta}$ and $\hat{\phi}$ can be expressed as 
\begin{align}
	\text{var}(\hat{\theta}) \ge \frac{\sigma^2}{2Ks^2}\frac{R}{PR-Q^2},\nonumber\\
	\text{var}(\hat{\phi})\ge \frac{\sigma^2}{2Ks^2}\frac{P}{PR-Q^2}.
\end{align}

Let $\gamma=\frac{2Ks^2}{\sigma^2}$ is the received SNR at the array. Then the closed-form expression of the CRLB can be written as
\begin{align}
	\text{var}(\hat{\theta}) &\ge \frac{1}{\gamma} \times \frac{d_2 \beta_2^2 + d_1 \alpha^2 R_1}{d_4 \alpha^{2p} \left[d_2 G + d_3 \sin^2\theta \cdot \alpha^2\right]}\label{CRLB_theta},  \\
	\text{var}(\hat{\phi}) &\ge \frac{1}{\gamma} \times \frac{d_2 \beta_1^2 + d_1 \alpha^2 P_1}{d_4 \alpha^{2p} \left[d_2 G + d_3 \sin^2\theta \cdot \alpha^2\right]}\label{CRLB_phi} ,
\end{align}
where $d_2$, $d_3$, $d_4$, $P_1$, $Q_1$, $R_1$ and $G$ are given in Eq. (\ref{d}).
\begin{figure*}[htbp]
	\centering
	\begin{align}\label{d}
		d_2=MNp^2,&~~~~
		d_3=d_1MNS_mS_n,~~~~
		d_4=d_1g_0^2,\nonumber\\
		P_1(\Delta_{\theta},\Delta_{\phi},\theta, \phi )&=\cos^2\theta\cos^2(\phi+\Delta_{\phi})MS_n
		+(\cos\theta\sin(\phi+\Delta_{\phi})\sin\Delta_{\theta}+\sin\theta\cos\Delta_{\theta})^2NS_m,\nonumber\\	
		Q_1(\Delta_{\theta},\Delta_{\phi},\theta, \phi )&=\sin\Delta_{\theta}\cos\Delta_{\theta}\sin^2\theta\cos(\phi+\Delta_{\phi})NS_m
		-\sin\theta\cos\theta\cdot\sin(\phi+\Delta_{\phi})\cos(\phi+\Delta_{\phi})(MS_n-NS_m\sin^2\Delta_{\theta}),\nonumber	\\
		R_1(\Delta_{\theta},\Delta_{\phi},\theta, \phi )&=\sin^2\theta[\sin^2(\phi+\Delta_{\phi})MS_n+\sin^2\Delta_{\theta}\cos^2(\phi+\Delta_{\phi})NS_m],\nonumber	\\
		G(\Delta_{\theta},\Delta_{\phi},\theta, \phi )&=\beta_1^2R_1+\beta_2^2P_1-2\beta_1\beta_2Q_1. 
	\end{align}
	\hrule 
\end{figure*}

Specially, for the array before rotation as Fig. \ref{fig:xyz}\subref{subfig:xyz-a}, i.e.  $\Delta_{\theta}= 0^o$ and $\Delta_{\phi}=0^o$, the CRLB of initial array is obtained by Eq. (\ref{theta0}) and Eq. (\ref{phi0}). 
\begin{figure*}  
	\centering
	\begin{align}\label{theta0}
		\text{var}(\hat{\theta}) &\ge \frac{1}{\gamma} \frac{ N p^{2} \cos^{2}\phi + \left( \frac{2\pi}{\lambda} \right)^{2} S_{n} \cos^{2}\varphi\sin^{2}\phi }{ g_{0}^{2} (\cos\varphi)^{2p} \left( \frac{2\pi}{\lambda} \right)^{2} N \left[ M p^{2} S_{n} \cos^{2}\theta + N p^{2} S_{m} \sin^{2}\theta \cos^{2}\phi + \left( \frac{2\pi}{\lambda} \right)^{2} S_{m} S_{n} \cos^2\varphi \right] }, \\
		\text{var}(\hat{\phi}) &\ge \frac{1}{\gamma} \frac{  M N p^{2} \sin^{2}\phi\cos^{2}\theta + \left( \frac{2\pi}{\lambda} \right)^{2} \cos^{2}\varphi\left( M S_{n} \cos^{2}\theta  \cos^{2}\phi+ N S_{m}  \sin^{2}\theta \right)  }{ g_{0}^{2} (\cos\varphi)^{2p} \left( \frac{2\pi}{\lambda} \right)^{2} M N \sin^{2}\theta \left[ M p^{2} S_{n} \cos^{2}\theta + N p^{2} S_{m} \sin^{2}\theta \cos^{2}\phi + \left( \frac{2\pi}{\lambda} \right)^{2} S_{m} S_{n} \cos^2\varphi \right] }.\label{phi0}
	\end{align}
	\hrule 
\end{figure*}

%
%
%
%
%
%

Observing the right-side of the above  inequalities and Fig. 2(b), if we rotate the antenna at  BS such that the emitter direct aligns with  or is parallel to the normal vector, then the above inequalities reduce to
\begin{align}
	\text{var}(\hat{\theta}) &\ge \frac{1}{\gamma} \times \frac{1}{\frac{2\pi}{\lambda}NS_mg_0^2},  \\
	\text{var}(\hat{\phi}) &\ge \frac{1}{\gamma} \times \frac{1}{\frac{2\pi}{\lambda}MS_ng_0^2\sin^2\theta} .
\end{align}

\section{Simulation Results}
In this section, we present simulation results to evaluate the performance of proposed RR-Root-MUSIC method and the corresponding CRLB as a performance benchmark. The parameters of simulation are set as: $N=M=6$, $P_t$= 20dBm, $\sigma^2 $= -100dBm, $\phi=90^\circ$, $r$ = 250m, $\lambda$ = 0.125, $A$ = $\lambda^2/4\pi$. In our simulation, all results are averaged over 2000 Monte Carlo realizations and RMSE is calculated to indicate the performance, which is given by
\begin{align}
	\text{RMSE} = \sqrt{\frac{1}{L}\sum_{l}^{L}(\theta_l-\theta)^2},
\end{align}	
where $L$ denotes the times of Monte Carlo experiments.

\begin{figure}
	\centering
	\subfloat[L = 1]{
		\includegraphics[width=0.5\textwidth]{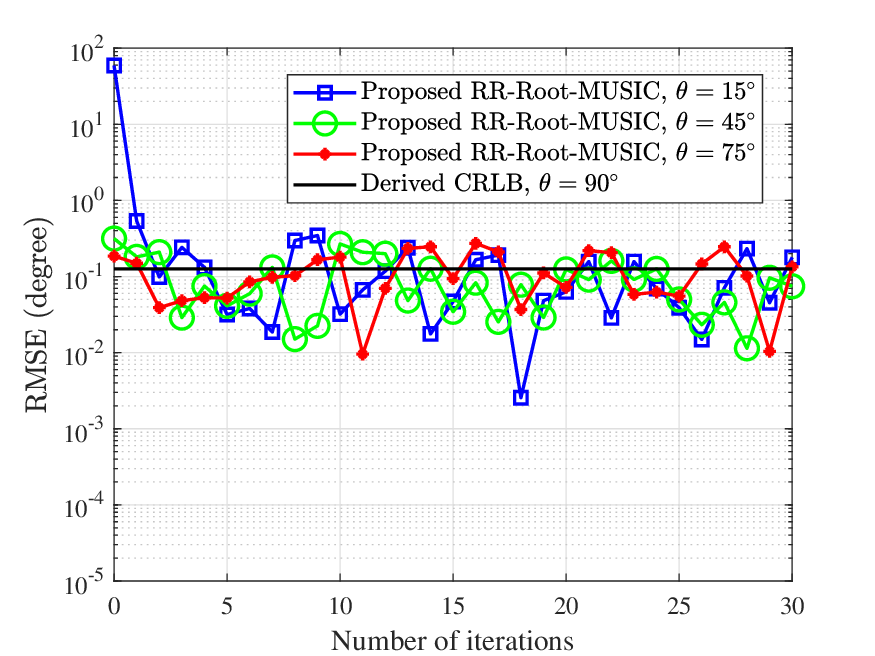}
		\label{fig:conver-curve-low_L1}
	}
	\hfill
	\subfloat[L = 1000]{
		\includegraphics[width=0.5\textwidth]{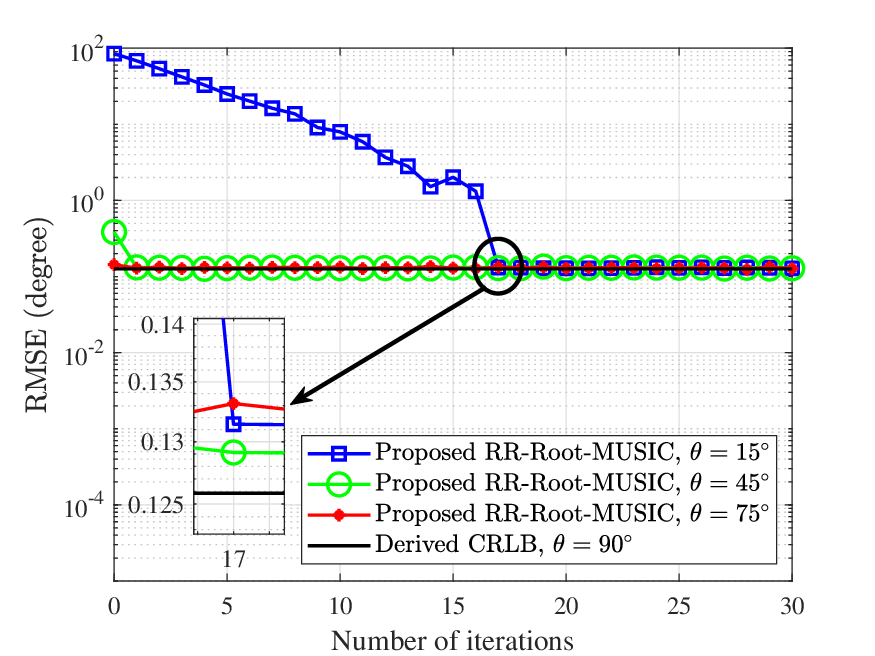}
		\label{fig:conver-curve-low_L1000}
	}
	\caption{Convergence curves of the proposed RR-Root-MUSIC method when SNR = -10 dB.}
	\label{fig:conver-curve-low}
\end{figure}

In Fig. \ref{fig:conver-curve-low}, the RMSE convergence curves of the proposed RR-Root-MUSIC method are illustrated for SNR = -10 dB and three different elevation angles:   $\theta=75^\circ$, $\theta=45^\circ$, $\theta=15^\circ$, with  CRLB as a performance benchmark, where $L=1$ in Fig. \ref{fig:conver-curve-low}\subref{fig:conver-curve-low_L1} and $L=1000$ in in Fig. \ref{fig:conver-curve-low}\subref{fig:conver-curve-low_L1000}. Obviously, given  a fixed value of SNR, the proposed method asymptotically converges to the corresponding error floor after about 13 or less rotations on average. As the angle of the emitter deviates away from the normal vector of the rotation array, the RMSE of the harvested performance gain  ranges from  one-order magnitude  to  three-order magnitude as  $\theta$ varies from $75^\circ$ to $15^\circ$. This means that the larger the angle difference between the incident direction of the emitter and the normal direction of array, the larger the achieved performance gain by the proposed rotation operation. When $\theta$ is close to $90^\circ$, the proposed method takes only one time to converge to the same error floor. In summary, if the emitter direction is far away from the normal vector, a significant RMSE performance gain may be achieved.

\begin{figure}
	\centering
	\subfloat[L = 1]{
		\includegraphics[width=0.5\textwidth]{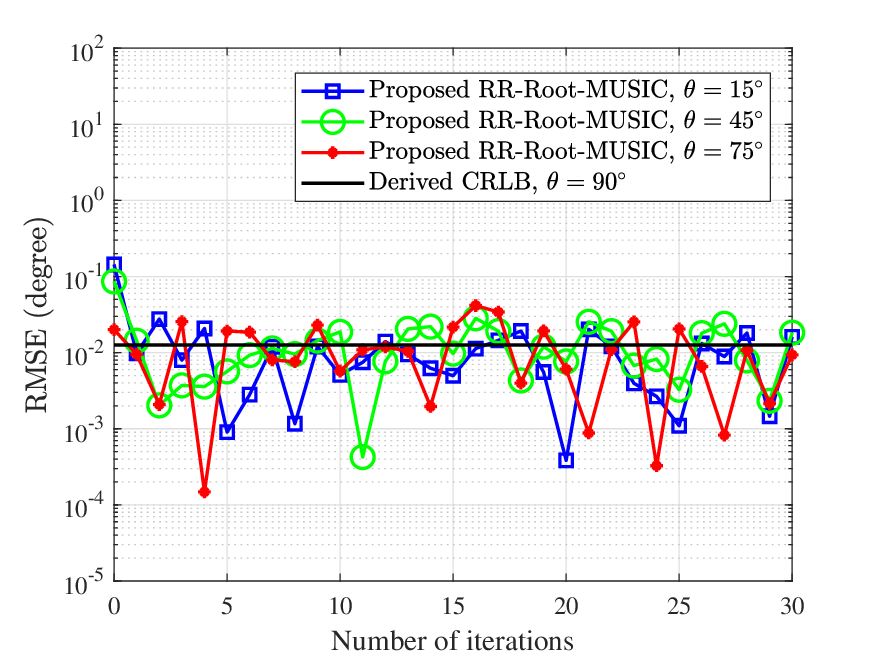}
		\label{fig:conver-curve-midd_L1}
	}
	\hfill
	\subfloat[L = 1000]{
		\includegraphics[width=0.5\textwidth]{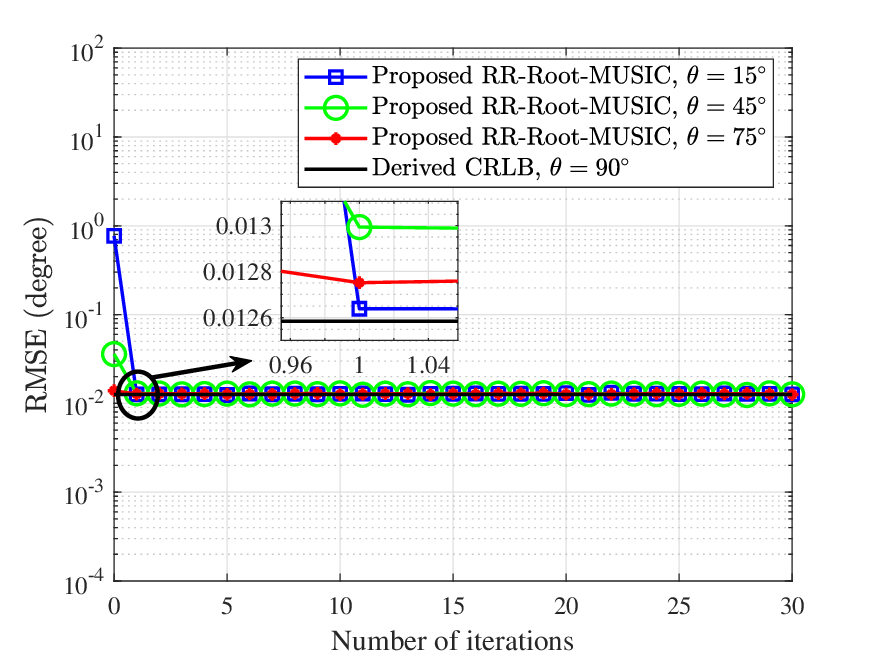}
		\label{fig:conver-curve-midd_L1000}
	}
	\caption{Convergence curves of the proposed RR-Root-MUSIC method when SNR =  10 dB.}
	\label{fig:conver-curve-midd}
\end{figure}

\begin{figure}
	\centering
	\subfloat[L = 1]{
		\includegraphics[width=0.5\textwidth]{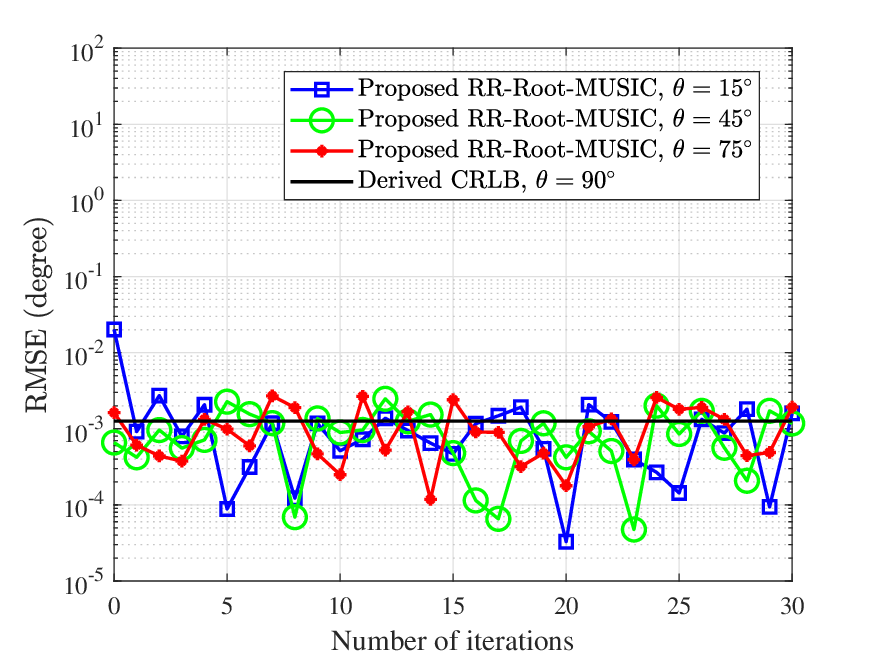}
		\label{fig:conver-curve-high_L1}
	}
	\hfill
	\subfloat[L = 1000]{
		\includegraphics[width=0.5\textwidth]{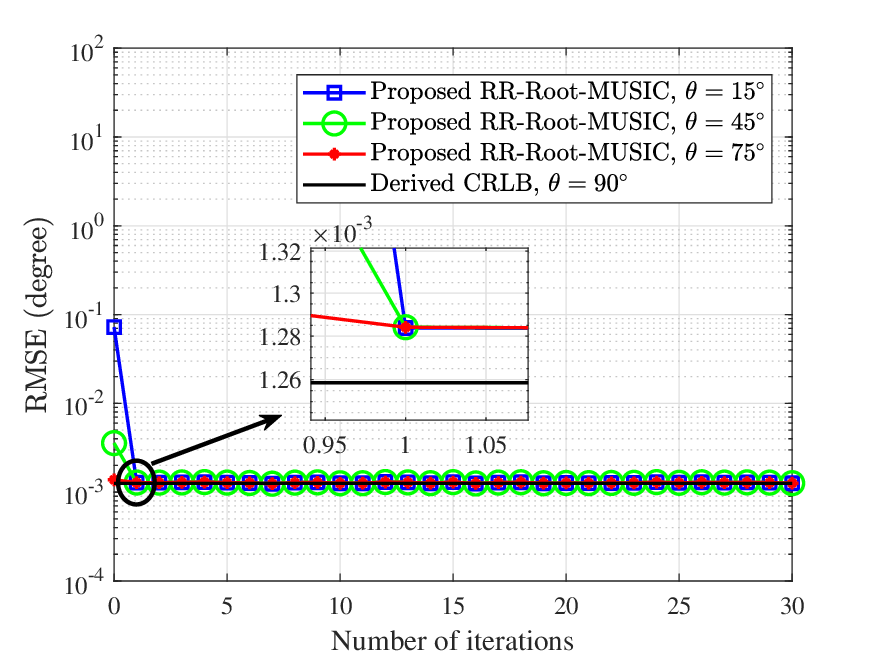}
		\label{fig:conver-curve-high_L1000}
	}
	\caption{Convergence curves of the proposed RR-Root-MUSIC method when SNR = 30 dB.}
	\label{fig:conver-curve-high}
\end{figure}
Similar to Fig. \ref{fig:conver-curve-low}, in Figs. \ref{fig:conver-curve-midd}$\sim $\ref{fig:conver-curve-high},  the same curves  are plotted for two typical  SNRs : \{10 dB\},  and \{30 dB\}.  Observing the three figures, it is very clear:  as SNR grows from -10 dB to 30 dB, the convergence rate is improved gradually, i.e. the required number of iterations has been reduced from 13 to 1 on average.  Similar to Figs. \ref{fig:conver-curve-low}, a significant RMSE performance gain is achieved when the angle difference $90^o-\theta$ between the target direction  and the boresight direction nears $90^o$, i.e, $\theta \rightarrow   0^o$ .


\begin{figure}
	\centering
	\includegraphics[width=1\linewidth]{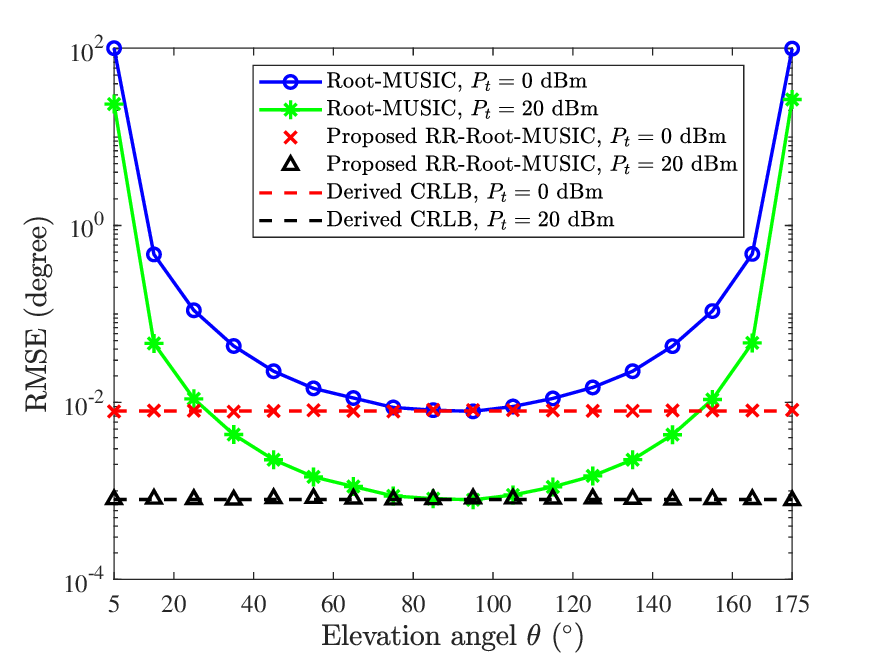}
	\caption{RMSE versus the elevation angles $\theta$ for proposed RR-Root-MUSIC method with transmit powers $P_t\in$ \{0 dBm, 20 dBm\}. }
	\label{fig:rmseptRAA}
\end{figure}
In Fig. \ref{fig:rmseptRAA}, the  RMSE curves of the proposed RR-Root-MUSIC method versus the elevation angle $\theta$ are shown for different transmitter powers $P_t\in$ \{0 dBm, 20 dBm\}, with CRLB as a performance benchmark. It is obvious that as $\theta \rightarrow   0^o$, the performance gain of the proposed RR-Root-MUSIC over conventional fixed Root-MUSIC  grows gradually. In extreme situation, i.e., $\theta \le 5 ^o$, the RMSE performance of the proposed method is more four-order  magnitude better than that of the conventional fixed Root-MUSIC. This further confirms  that a rotatable array may effectively mitigate the pattern-induced performance loss.

\begin{figure}
	\centering
	\subfloat[UAV motion path]{
	\includegraphics[width=0.5\textwidth]{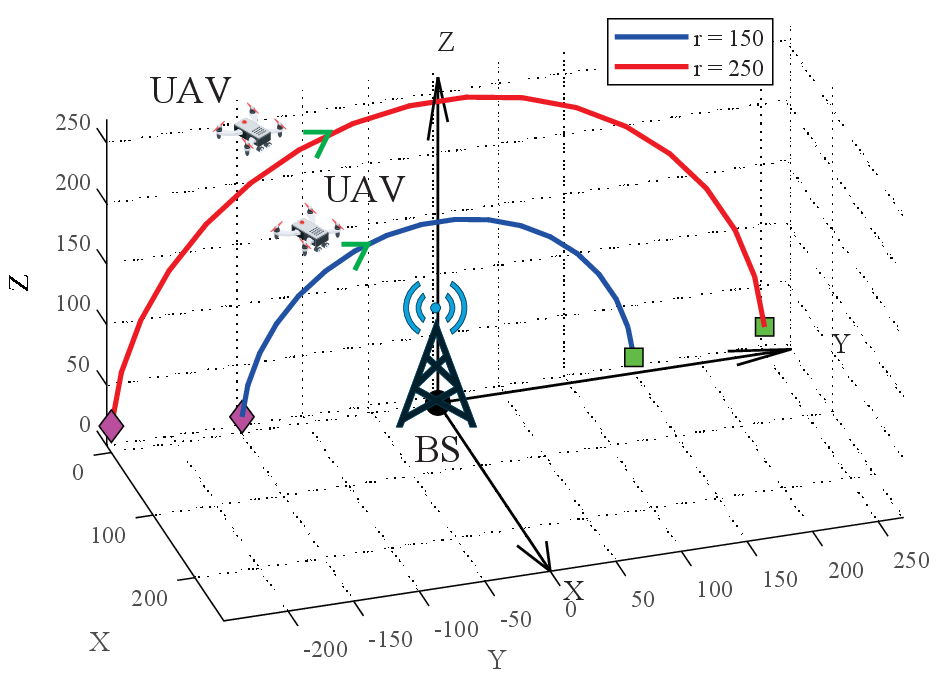}
	\label{fig:uav path}
	}
	\hfill
    \subfloat[RMSE versus the motion path of UAV]{
	\includegraphics[width=0.5\textwidth]{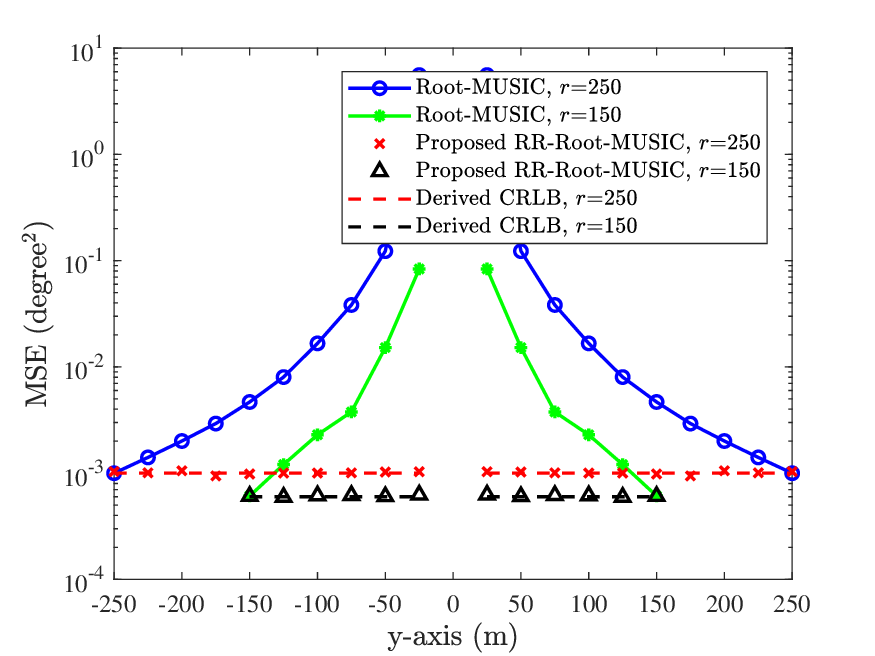}
	\label{fig:MSE uav}
}
	\caption{The motion path of the UAV around BS, and RMSE of proposed RR-Root-MUSIC method versus the motion path of UAV.}
	\label{fig:combined}
\end{figure}

To observe the influence of the proposed rotation method in low-altitude UAV flying, Fig. \ref{fig:combined}\subref{fig:uav path} shows the motion path of the UAV around BS.  Fig. \ref{fig:combined}\subref{fig:MSE uav} illustrates the corresponding RMSE performance curves along semi-circular trajectory of UAV for the proposed RR-Root-MUSIC and the traditional fixed MUSIC method. From Fig. \ref{fig:combined}\subref{fig:uav path}, the UAV flies form (0, -250, 0) and (0, -150, 0) to (0, 250, 0) and (0, 150, 0) with radii $r=250$ and $r=150$, respectively. Compared with the traditional Root-MUSIC method, the proposed RR-Root-MUSIC method makes a dramatic performance gain over the conventional fixed Root-MUSIC when the position of the UAV is on top of BS, which is usually a blind coverage area. 
Additionally, due to the use of rotatable array, the sensing performance keeps stable during the flight process of UAV.  Thus, the proposed rotatable array system and method will offer an excellent new solution to the future low-latitude communication network.


\section{Conclusions} 
\noindent

In this paper, to solve  the low-altitude coverage problem of MIMO transceiver with  directional antenna elements in practice, a rotatable array system was established with each element being the same  directional pattern. Here, via a sensing and rotation iterative loop, the RR-Root-MUSIC was proposed  to fully  harvest the antenna array gain to make a dramatic performance enhancement, especially for the scenario that the emitter direction is far away from the normal vector of array. Compared with classical Root-MUSIC without rotation, the proposed method may achieve up to three-order magnitude improvement or more at the elevation  angle $\theta \le 15^o$ in terms of RMSE performance. Moreover, the proposed RR-Root-MUSIC converges fast and provides a full coverage on such an area on top of BS,  which cannot be covered by the conventional fixed antenna array system like 5G or 4G.  Due to the above advantages of rotation-array system, it will become an extremely potential technique for the future low-altitude wireless network to completely break the bottleneck of low-latitude coverage faced by  the current mobile communication like 5G.

%

\end{document}